# The effect of side chain length on the mesomorphic properties of ferroelectric nematogens

Natalia Podoliak,[1*] Martin Cigl,[1] Pavlo Golub,[1] Anej Sterle,[2,3] Marta Lavrič,[2] Nerea Sebastian,[2] Aitor Erkoreka,[4] Josu Martinez-Perdiguero,[4] Alenka Mertelj,[2] George Cordoyiannis,[2] and Vladimíra Novotná[1]

*[1] Institute of Physics of the Czech Academy of Sciences, Na Slovance 1999/2, 182 00 Prague 8, Czech Republic*
*[2] Jožef Stefan Institute, Jamova 39, 1000 Ljubljana, Slovenia*
*[3] University of Ljubljana, Faculty of Mathematics and Physics, 1000 Ljubljana, Slovenia*
*[4] Department of Physics, Faculty of Science and Technology, University of the Basque Country UPV/EHU, Bilbao, Spain*

Corresponding author*: podoliak@fzu.cz

Key words: liquid crystals, ferroelectric nematics, SHG measurements, dielectric spectroscopy

Abstract
The discovery of ferroelectric nematic phase opened new directions in the field of soft matter chemistry and physics. In this paper, we have modified a previously reported ferroelectric nematogen based on molecular structure with dimethylamino-terminated part. We have prepared two new homologues by prolonging a side chain and investigated the effect of its length on the mesomorphic properties. While previously reported homologues exhibited a direct phase transition from the isotropic (Iso) to the ferroelectric nematic phase ($N_F$), for prolonged alkyl chain there is a narrow nematic phase (N) in between. For new compounds, we have established their mesomorphic and material properties. We have confirmed ferroelectricity of the $N_F$ phase by the polarization, SHG signal and permittivity. Finally, we have compared these parameters with respect to the side-chain length and discussed new effects, discovered for prolonged homologues due to the presence of an additional N phase.

## Introduction

The discovery of ferroelectricity of the nematic phase can be considered as a significant achievement not only in the field of liquid crystals, but within a broader soft matter concept. Since 2017, when two groups introduced the first two mesogens revealing ferroelectricity in a nematic phase [1-2], research activity has intensively grown in the field of highly polar molecules [3-20]. Ferroelectric nematic phase ($N_F$), as a polar version of traditional apolar nematic one (N), possesses unprecedented and remarkable properties: unusually high susceptibility to applied electric field, huge electro-optical and piezoelectric response [21-23]. Preserving the fluidity of liquid crystals, the $N_F$ phase exhibits extraordinarily high polarisation, which is comparable with solid ferroelectrics, [22] and its non-linear optical properties appear to be very promising for applications [23-24].

In the N phase, the molecular director *n* is sign-invariant and the phase is paraelectric. On the contrary, the ferroelectric nematic phase, $N_F$, represents a 3D liquid with a macroscopic polarization. The necessary condition for the $N_F$ phase realisation is overall molecular dipole magnitude vs. variation of dipole across the structure [6]. However, there is a delicate balance between molecular dipole moment and overall molecular shape anisotropy to promote polar order [6-9, 25-26]. In addition to these two nematic phases, an antiferroelectric phase appears between the N and $N_F$ for certain molecular structures, referred to as $SmZ_A$ phase [27-29]. It has been established by a resonant X-ray study that in the $SmZ_A$ phase polar nematic layers with alternating directions of dipole moment are formed [28].

For the $N_F$ phase, interactions with the cell surfaces differ from ordinary nematics due to their polar character. Various structures and domains can be created in relation to the anchoring conditions in the cell interior [30-33]. For liquid crystal (LC) research, two basic cell geometries are commonly utilised: planar/homogeneous (HG) arrangement, with molecules parallel to the confining surfaces, or a homeotropic (HT) geometry with perpendicular molecular orientation at the cell surfaces. When the upper and bottom surfaces in the HG cells are covered with antiparallelly rubbed surfactant (the HG-A cell), the transition into the $N_F$ phase is accompanied with the creation of director twist within the sample thickness. Domains of two opposite senses of twist are typically observed for such cell geometry. Neighbouring twisted domains are separated by defects, which were explained to be $2\pi$ disclinations [31].

Nevertheless, the relationship between the molecular structure and existence/stability of the $N_F$ phase is still not fully comprehended. Nowadays, intensive design of new ferroelectric nematogens indicates that even a small change in the molecular structure can drastically influence mesogenic properties [6-9, 26]. Not only dipole-dipole interactions, but also mutual intermolecular interactions and the overall shape are critical for the appearance of ferroelectricity in nematic phase. Scientists intend to design new ferroelectric nematogens with the aim to maximize their stability at ambient temperatures [34-38] due to growing demand for applications. Unfortunately, vast majority of up-to-date molecular structures are still not ideal for practical applications, as they are mostly mesogenic at high temperatures, are not enough chemically or thermally stable. In another respect, preparation of eutectic mixtures does not allow scientists to understand all molecular aspects.

In the previous research, we designed an original molecular structure with diethylamino moiety as an effective electron donating group (Fig. 1) and a nitro group at the opposite side of the molecule [37]. We prepared homologues with n=1-6 (Fig. 1) and demonstrated the presence of a direct phase transition from the isotropic (Iso) to the ferroelectric nematic phase ($N_F$). With the prolongation of the side chain, we succeeded in lowering the phase transition temperatures and longer homologues (n=5,6) revealed stable $N_F$ phase persisting at ambient temperatures. This motivated us to pursue further side chain prolongation. In this contribution, we have prepared two additional homologues with prolonged side chain (n=7 and n=8) and investigated their mesomorphic properties by several experimental techniques including observation of textures, differential scanning and AC calorimetry, polarization and SHG measurements as well as determination of second order nonlinear optical coefficients, optical birefringence measurements and dielectric spectroscopy. In comparison with previously studied materials, new homologues possess intermediate narrow N phase between the Iso and $N_F$ ones, the temperature interval of which increased with the chain length prolongation. Longer side chain and the presence of N phase influenced the molecular packing and alignment of the investigated compounds, which reflected in materials characteristics that are thoroughly described in this paper.

## Experimental

### Materials

The chemical formula of the studied materials is presented in Fig. 1. New compounds were synthesised according to the procedures described previously for **NFn** compounds with n=1-6 [37]. Synthetic pathways and characterization of **NF7** and **NF8** homologues are detailed in Supplementary file (SI). Both homologues are proved to exhibit the phase sequence Iso-N-$N_F$ on cooling from the isotropic (Iso) phase. Detailed analysis will be described below.

NF7 n=7
NF8 n=8

Fig. 1. Chemical formula of the studied molecules **NF7** and **NF8**.

### Experimental methods and apparatus

For the observation and acquisition of textures and their modifications with temperature, we utilised a polarized light microscope Nikon Eclipse E600, equipped with a heating-cooling stage Linkam LTS350E, controlled by a temperature controller Linkam TMS 94 (Linkam, Tadworth, UK), with an accuracy of temperature stabilization ±0.1 K, and a photo camera Canon EOS 700D. For electro-optical studies, commercial cells (WAT PPW company, Poland) of various geometries were purchased: homogeneous (HG) with the rubbing direction of the surfactant layer parallel (HG-P) or antiparallel (HG-A) at the opposite surfaces.

For the phase transition temperatures and enthalpy changes determination, two types of calorimeters were used. Differential scanning calorimetry (DSC) was performed using a Perkin Elmer calorimeter. For such measurements, the samples of 2-5 mg were hermetically sealed into aluminium pans and placed inside the calorimeter chamber inflated with nitrogen. The measurements were performed during heating and cooling cycles at a rate of 10 K/min. For narrow mesophase temperature regions, a home-made experimental setup built at Jožef Stefan Institute, with thermal stability better than 0.1 mK, was utilised. This apparatus allows us to perform very slow cooling and heating runs down to a few tenths of K per hour. Technical details about the operation of the AC calorimetric method can be found in our SI file, as well as in literature [39].

Dielectric spectroscopy studies were conducted with an impedance analyser Solartron. For these studies, 7-μm plane-parallel cells with and without surfactant layers were used. The sample temperature was stabilised within ± 0.1 K during the frequency sweeps (1 Hz ÷ 1 MHz). From the resistance and capacity of the sample, real and imaginary parts of the complex permittivity, $\varepsilon^*(f) = \varepsilon' - i\varepsilon''$, were measured and the results were fitted to the Cole-Cole formula to obtain the relaxation frequency, $f_r$, and the dielectric strength, $\Delta\varepsilon$. For details, see Supplementary information (SI). Polarization, $P$, measurements were conducted in the electric field of triangular profile, of frequency 10 Hz and a magnitude of 10 V/μm, produced by a generator Agilent 33210A. For these measurements, we utilised 25-μm home-made cells without surfactant layer. At stabilised temperature, a switching current was detected and integrated to obtain $P(T)$ values.

For optical birefringence evaluation, a set-up based on photoelastic modulator PEM-100 (Hinds instruments) (PEM) was used [40]. For these measurements, commercial 1.6-μm parallelly rubbed planar cells (HG-P) were utilized. The samples were placed into Instec heating stage equipped with Instec STC 200 temperature controller. A green light ($\lambda$=550 nm) beam passed through a polariser positioned at 45°, subsequently through PEM, the sample, analyser positioned at -45° and a photodetector. Birefringence of the sample was calculated from the measured retardation using the equation:

$$\Delta n = \delta \cdot \lambda / (2\pi d), \qquad (1)$$

where $\delta$ is the retardation and $d$ is the thickness of the sample.

The second harmonic generation (SHG) imaging is performed using a custom-built sample-scanning microscope. For details and scheme, see SI (Fig. S1). The laser source is an Erbium-doped fibre laser (C-Fiber A 780, MenloSystems) generating 785 nm, 95 fs pulses at a 100 MHz repetition rate. The average power was adjusted using an ND filter to 30 mW on the sample. To determine the second order dielectric susceptibility coefficient $d_{33}$, we employ Maker's fringes method in a second setup consisting of a Q-switched pulsed Nd:YAG laser with an optical parametric oscillator module operating at λ=1574 nm. This excitation wavelength was chosen, so that the generated SHG signal remain well outside the electronic absorption spectrum of the material, thereby minimizing linear absorption and thermal effects while enabling reliable measurement of the intrinsic nonlinear optical response. The beam is collimated, linearly polarized and the spot size is limited using a diaphragm. The intensity of the emitted second harmonic light (λ=787 nm) is measured by a photomultiplier after going through an analyser. For these measurements, custom-made wedge cells are fabricated with top and down electrode gap of 2 mm. See SI for a detailed description. Wedge angle was calibrated via interferometric methods before filling and determined to be $tan\theta = 0.0017$ and $tan\theta = 0.0026$, for the cells employed for **NF7** and **NF8** respectively.

# Results

## DSC measurements

To obtain phase transition temperatures and enthalpy changes, compounds **NF7** and **NF8** were studied by DSC thermal cycles, and the results are collected in Table 1, where the temperatures for **NF6** compound taken from [37] are added for comparison. We established the melting points (m.p.) from the first heating runs, during which we observed transitions directly to the isotropic (Iso) phase. On the subsequent cooling runs from the Iso phase down to -25°C, we detected two peaks corresponding to the Iso-N and N-$N_F$ phase transitions at temperatures $T_{iso}$ and $T_c$, respectively. For both compounds, a glass transition was observed. For **NF8** compound, an exothermic peak corresponding to cold crystallization appeared above the glassy transition during the second heating run, which corresponds to the crystallization upon heating from the amorphous state into more ordered crystalline one [41]. In Fig. 1, we compare the phase transition temperatures for all **NFn** homologues, taking the data from the previous studies for n=1-6, see [37]. For both **NF7** and **NF8** compounds, the paraelectric nematic phase exists only in a narrow temperature interval of 2-3 K. From Fig. 1 it is evident that the transition temperature to ferroelectric nematic phase decreases with the prolongation of the side chain length (increase in n number).

**Table 1.** The temperatures and enthalpy changes, obtained from DSC measurements. The melting points, m.p., were taken from the first heating runs. The phase transition from the isotropic phase (Iso) to the first mesophase, $T_{iso}$, the temperature of the N-$N_F$ phase transition, $T_{tr}$, and the crystallization temperature, $T_{cr}$, were established from the second cooling runs. Temperatures are in °C, and enthalpy changes, ΔH, in kJmol $^{-1}$, are in brackets at the corresponding temperatures. $T_g$ is the temperature of the glass transition and $\Delta c_p$ change in heat capacity is in kJmol$^{-1}$K$^{-1}$. Symbol – means the absence of the phase and/or the peak on DSC curves. A symbol * at the value of $T_{cr}$ means cold crystallization appearing upon a subsequent heating run [41].

| | m.p. [ΔH] | $T_g$ [$\Delta c_p$] | $T_{cr}$ [ΔH] | | | $T_{tr}$ [ΔH] | | $T_{iso}$ [ΔH] |
|---|---|---|---|---|---|---|---|---|
| **NF6** | 104 [+26.1] | 4 [0.14] | - | | | - | $N_F$ | 65 [-1.69] |
| **NF7** | 92 [+31.6] | -1 [0.22] | - | $N_F$ | | 58 [-1.17] | N | 60 [-0.26] |
| **NF8** | 99 [+30.7] | -5 [0.22] | 58* [-22.8] | $N_F$ | | 53 [-0.76] | N | 56 [-0.16] |

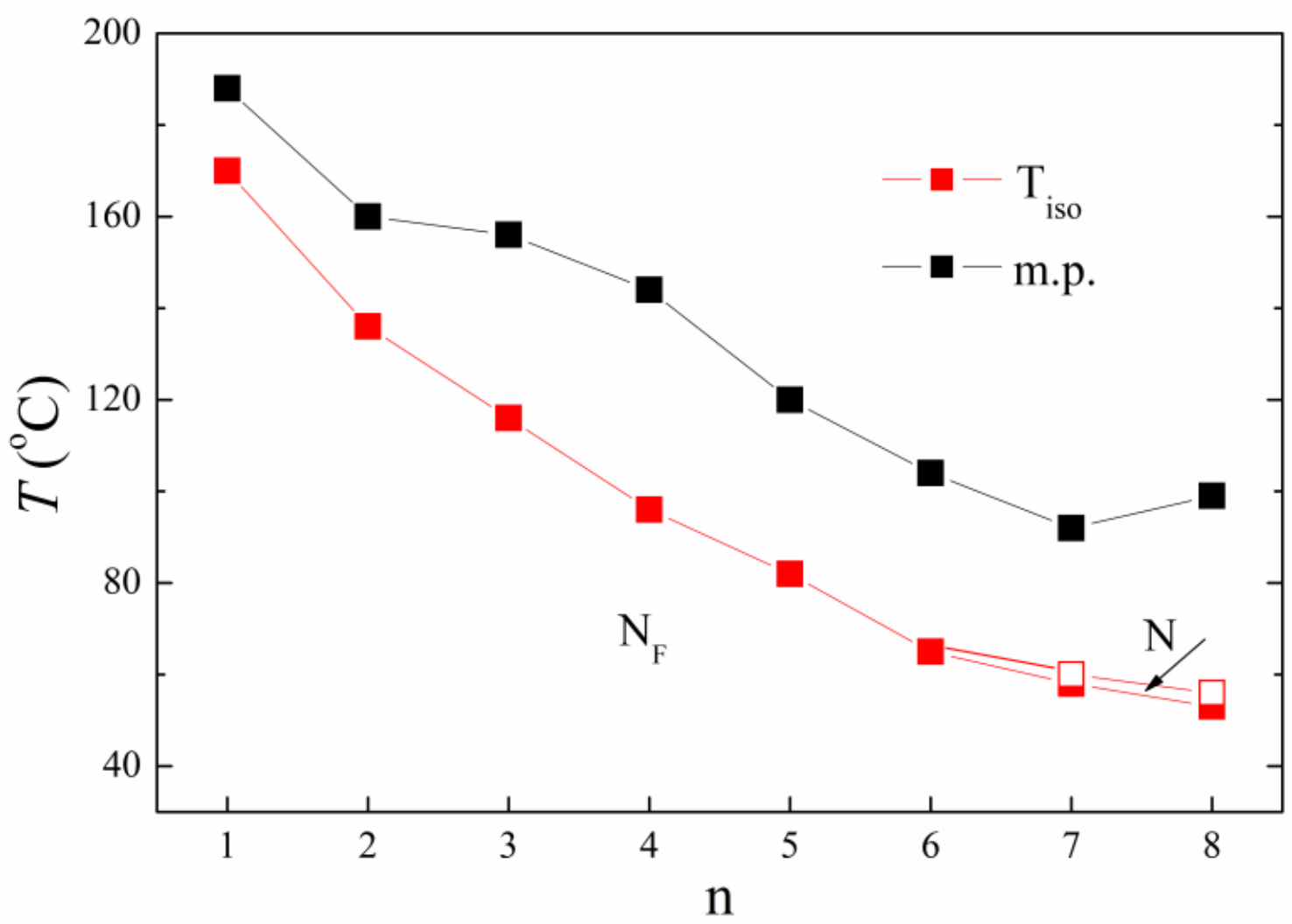

Fig. 1. Dependence of the phase transition temperatures on the n number, connected with the length of side chain, for all homologues **NFn**. Temperatures for the previously published homologues with n=1-6 are taken from Ref. [37].

**High resolution AC calorimetry measurements**

To precisely establish the phase transition behaviour, the temperature profile of specific heat capacity, $c_p(T)$ for sample **NF8** was measured using AC calorimetry. We focused on the vicinity of the N-$N_F$ phase transition. The $c_p(T)$ profiles on cooling and heating are presented in Fig. 2.

First, the sample was heated above the melting temperature to 105 °C, and the measurement was made upon cooling to room temperature with a scanning rate 150 mK/h (blue colour dots in Fig. 2), as this rate is sufficiently slow for observing phase transitions with small enthalpy content, in order to observe whether the mesophase $SmZ_A$ is present [42]. After first cooling to crystal phase, we heated the sample to isotropic phase and cooling was repeated down to ferroelectric nematic phase and then heated back to isotropic phase (red colour dots in Fig. 2). After melting to isotropic phase and keeping it above crystallization, two anomalies reproducible upon heating and cooling were observed, attributed to the Iso-N and N-$N_F$ phase transitions. Sharp positive peaks in phase behaviour indicate the first-order nature of both phase transitions as in the case of RM734 and DIO [42-44]. However, in contrast to RM734 and DIO, there is no indication for an intermediate $SmZ_A$ phase in the case of **NF8**; the nematic range is also narrow, $\Delta T_N = 2.5 \pm 0.1$ K. Note, that a shift in phase transitions temperatures obtained by AC calorimetry – compared to DSC – is reasonably expected, due to the three orders of magnitude smaller scanning rates in the former. The very fast rates of DSC – although useful to obtain the phase behavior at the initial stage of sample characterization – do not allow for sample thermal equilibrium and, especially for first-order transitions, they result in increased supercooling effects.

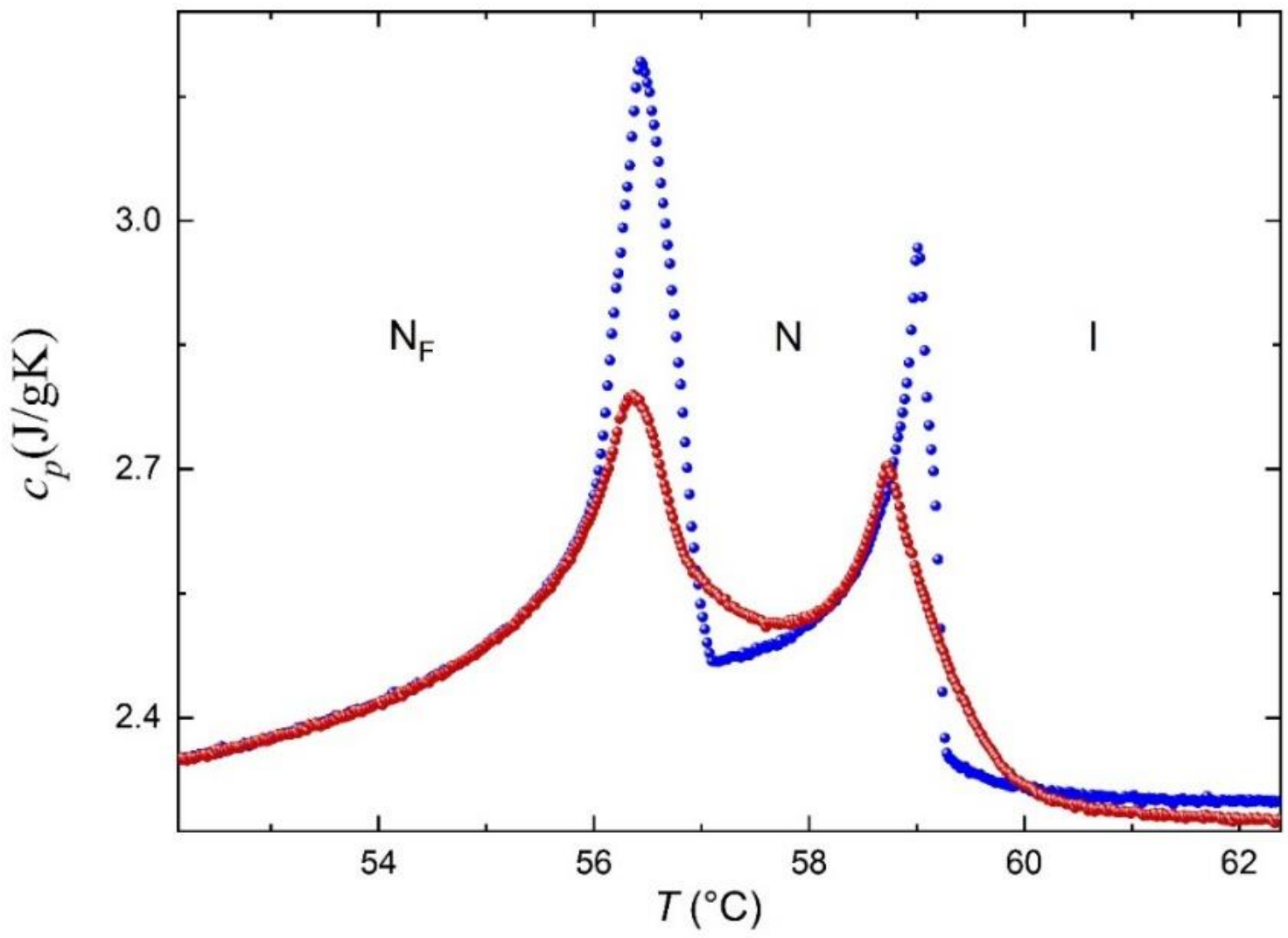

Fig. 2. Heat capacity ($c_p$) as a function of temperature for **NF8** upon cooling (blue dots) and heating (red dots) from nematic ferroelectric phase measured by AC high resolution calorimetry.

**Texture observations**

During observations in polarizing microscope (POM) on cooling from the isotropic phase, we observed characteristic nematic textures in all geometries. Nevertheless, the nematic phase in a cell without surfactant preferred homeotropic alignment and thus the textures were dark, only small defects were present and allowed us to distinguish between the isotropic and the nematic phases. In commercial cells with implemented molecular alignment, nematic textures appeared to be low-birefringent.

In $N_F$ phase, in commercial HG-A cells with an antiparallel rubbing, twisted domains with borderlines oriented perpendicularly to the rubbing direction dominated, similarly as it was observed for shorter **NFn** homologues [31]. Nevertheless, contrary to the previously studied homologues, for **NF7** and **NF8** we observed another type of domains oriented along the rubbing direction, which is demonstrated for **NF8** in a 5-μm HG-A cell in Fig. 3. The borderlines of these domains are presumably electrically charged, and their origins and structure are a topic of a separate study. Additional photos acquired during the microscope observations are presented in SI (Figs. S2-S4).

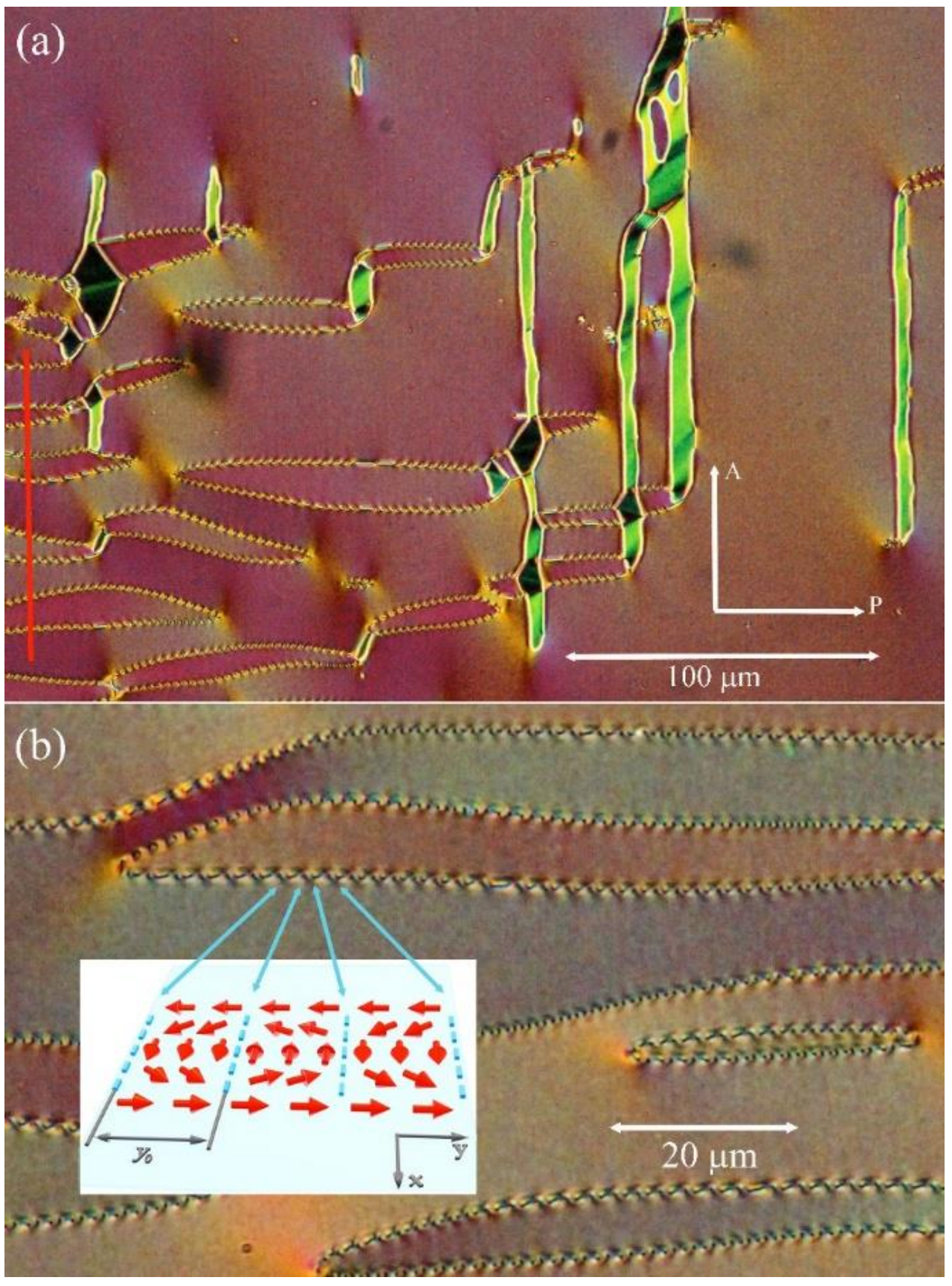

Fig. 3. Texture for **NF8** in the $N_F$ phase in a 5-μm HG-A cell. Figure (a) presents a large area of the cell with two types of domains described in the text, the rubbing direction is marked with red line; in (b) twisted domains with borderlines perpendicular to the rubbing direction are presented in an enlarged view, with point-like defects decorating disclination lines, which separate domains with the opposite sense of twist.

**Dielectric spectroscopy and polarisation**

To establish molecular dynamics, we have performed dielectric spectroscopy in a broad range of frequencies 1 Hz-1 MHz. In the literature, there are multiple controversial points regarding the interpretation of the results of dielectric spectroscopy for ferroelectric nematogens [45-49]. Thus, Clark et al. [45] attributed measured large dielectric strength values not to the bulk material properties but rather to the high capacitive insulating interfacial layers. They proposed that these layers serve as the principal energy storing mechanism of the Goldstone mode in this case and called it the polarization-capacitance Goldstone (PCG) mode. On the other hand, Vaupotič et al. proposed that the capacitance of the surface layers is comparable to the one of the bulk material, and the overall circuit must be considered.

To verify the role of surfactant layer, we have utilised two kinds of cells. Both compounds were measured in a 7-μm cell without surfactant, and in a commercial HG-A cell (with a surfactant layer) of the same thickness. During the measurements, we applied a low measuring field of 0.01 V/μm. The original data are presented as three-dimensional plots of the real, $\varepsilon$', and imaginary, $\varepsilon$'', parts of the complex permittivity in Figs. S5-S7. For both materials, we detected a very weak mode in the Iso phase, which moderately and continuously grew through the N phase, and a jump-up for orders of magnitude when entering the $N_F$ phase. The mode could be possibly attributed to the director fluctuations, the so-called phason mode. The relaxation frequencies of the mode decreased with the temperature lowering in the Iso phase. An abrupt jump-down at the Iso-N phase transitions was observed. The mode softened when approaching the N-$N_F$ phase transition and the relaxation frequency continued slightly decreasing with lowering temperature within the $N_F$ phase, which could be attributed to the increased rotational viscosity of the materials.

We fitted the experimental data to the generalized Cole-Cole equation (Eq. S1 in SI) to obtain the values of the dielectric strengths, $\Delta\varepsilon$, and the relaxation frequencies, $f_r$. The temperature dependences of fitted $\Delta\varepsilon$ and $f_r$ values are presented in Fig. 4 for **NF7** measured in both a home-made cell (full symbols) and a commercial HG-A cell (empty symbols). While the values measured in Iso phase are comparable for both home-made and commercial cells, there is a clear difference in the strength of the mode and the shift of the relaxation frequencies in the $N_F$ phase. The values of $\Delta\varepsilon$ are an order of magnitude higher for the cell without surfactant than those for the commercial cell, reaching up to 8000 in the cell without surfactant in comparison with 1000 in the cell with surfactant layer. We analysed the temperature dependences of the relaxation frequencies, $f_r$, see Fig. S8 in SI for **NF7**. Fitting to Arrhenius equation supplied us with activation energy, $E_a$, values. Contrary to classical N phase, we obtained similar activation energies in the Iso and $N_F$ phases (about 120±10 kJ/mol for **NF7** and 108±8 kJ/mol for **NF8**). For evaluation process we excluded temperature intervals near the phase transition temperatures due to pre-transitional fluctuations. Narrow temperature interval of the N phase did not allow us to establish $E_a$ in this phase.

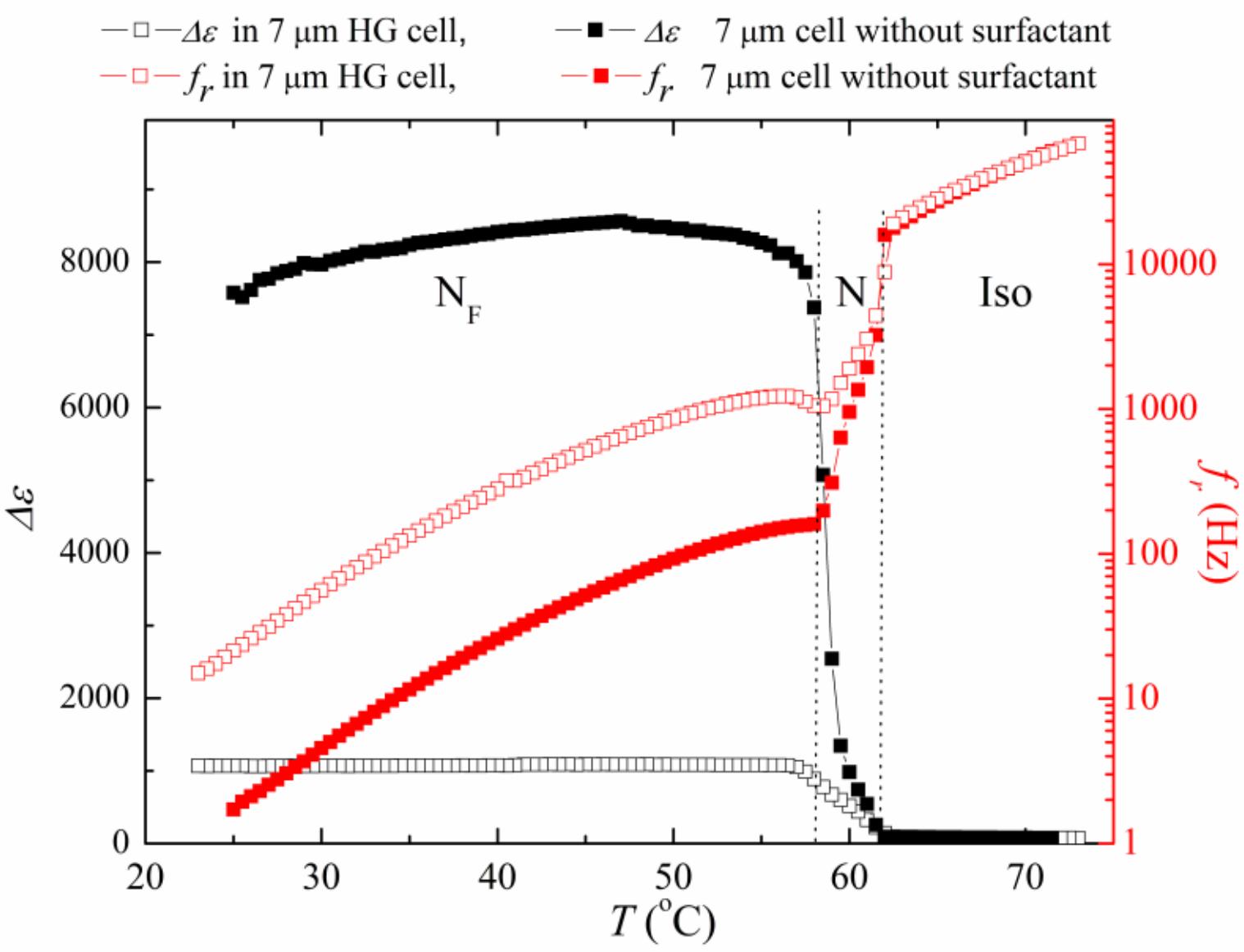

Fig. 4. The results of fitting of the dielectric spectroscopy data for **NF7**, measured for two plane-parallel cells of the same thickness, 7 μm, with and without surfactant layers applied onto the ITO electrodes: the dielectric strengths, $\Delta\varepsilon$, (black symbols), and the relaxation frequencies, $f_r$, (red symbols). Full symbols are the results for the cell without surfactant and empty symbols are the results for the cell with surfactant layer.

An appearance of a local minimum in the relaxation frequency values at the N-$N_F$ phase transition indicates the presence of an additional mode. To analyse this mode, we applied a bias electric field at T=55°C, close to the phase transition temperature. In Fig. 5, we demonstrate 3-dimensional graphs of the real and the imaginary parts of permittivity in frequency sweeps under an electric bias field changing from 0 to 0.48 V/μm. With increasing bias values, the frequency of the strong phason mode first remained practically unchanged (up to about 0.12 V/μm) and then became lowering. The strength of the mode did not change till about 0.2 V/μm bias and then started continuously decreasing with the bias increase. For the bias value of about 0.3 V/μm, the strong phason mode was suppressed enough and two relaxation processes (marked with arrows in Fig. 5(a)), one with low relaxation frequency around 10 Hz, and the other with higher relaxation frequency up to 1000 Hz, became apparent. The dielectric strength and relaxation frequency values for both relaxation modes, obtained by fitting of the experimental results to the Cole-Cole formula (Eq. S1 in SI) are presented in Fig. S9. The strength of the slower process decreased with increasing bias and the strength of the faster mode remained almost constant. While the relaxation frequency of the slower mode is practically independent of the bias field, the frequency of the faster relaxation process slightly increases with bias. This behaviour is in accordance with the one observed by Vaupotič et al. [47].

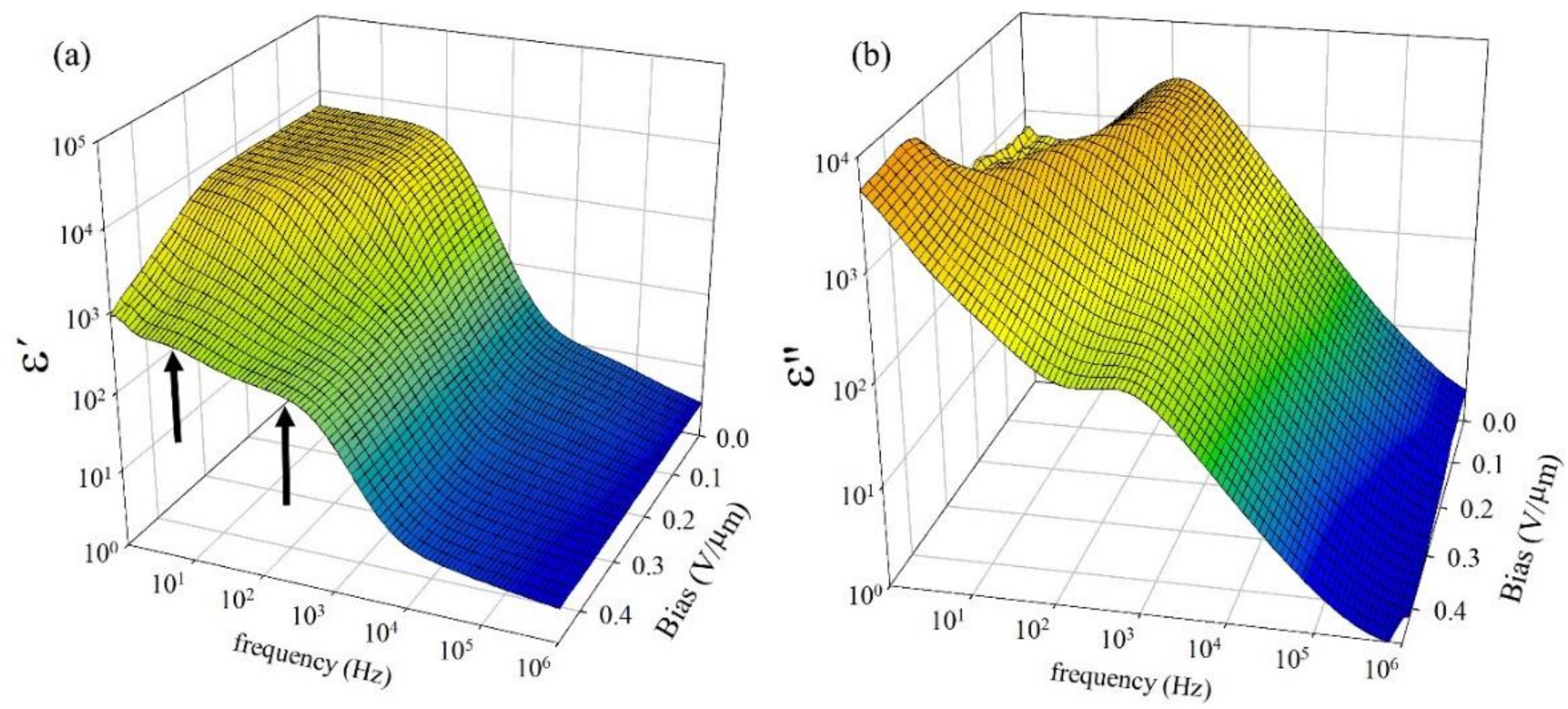

Fig. 5. (a) The real part of the complex permittivity, ε'; and (b) the imaginary part of the permittivity, ε'', in dependence on the applied bias field for **NF7** compound measured at T=55°C. Dielectric spectroscopy was performed in a 5-μm home-made cell without surfactant layers.

We investigated switching properties in the $N_F$ phase and established the values of polarization, *P*. In a similar manner for both compounds, a slight jump in *P*(*T*) values was observed at the N-$N_F$ transition, corresponding to the weakly first order nature of this transition, followed by a continuous growth with decreasing temperature throughout the $N_F$ phase. The saturation values reached up to 4.5 μC/cm$^2$. The temperature dependences of polarisation for both studied compounds together with NF6 homologue, which was added for comparison, are presented in Fig. 6. In the inset, one can see the switching current profile for **NF8** compound, measured at 48°C.

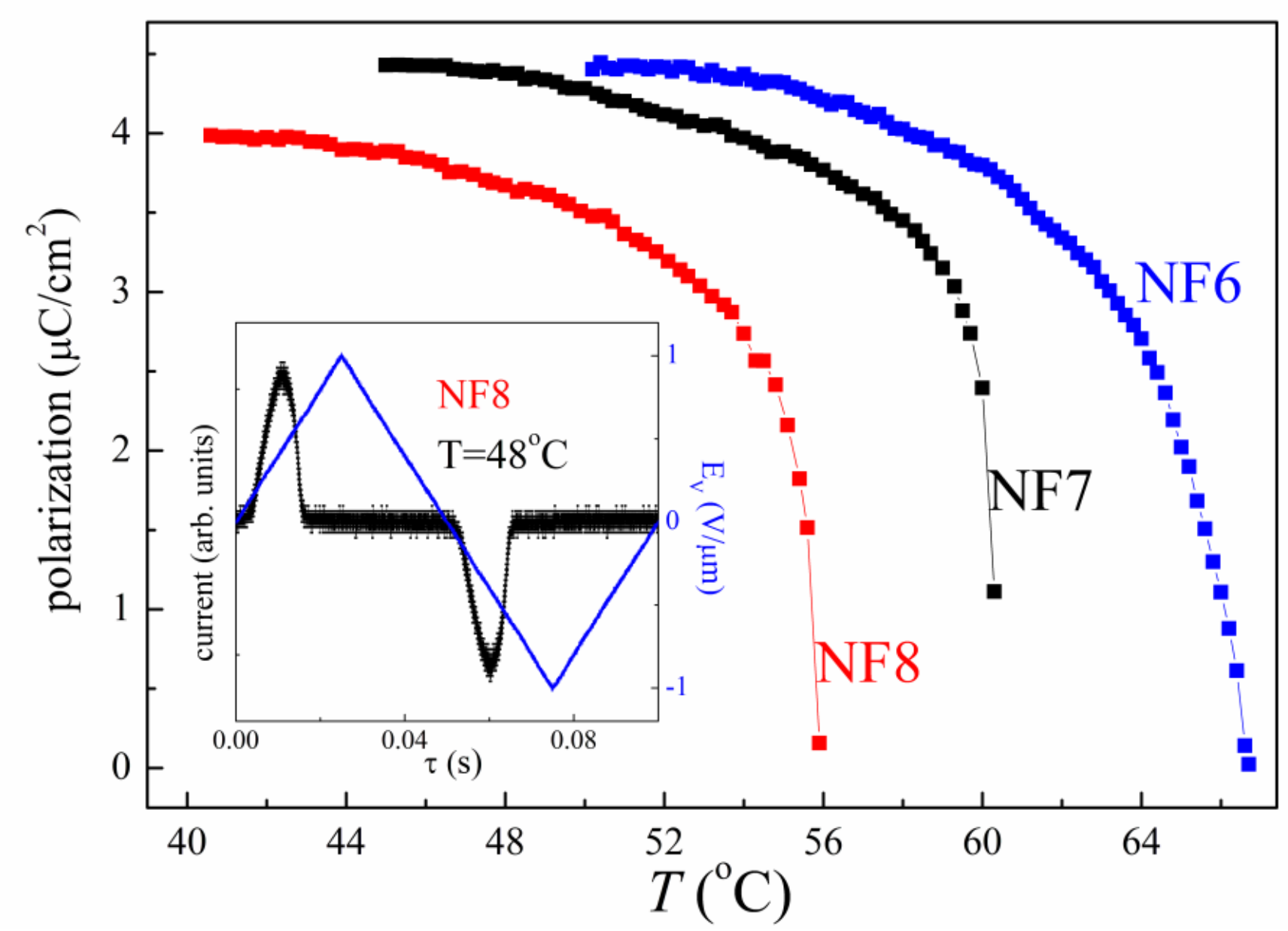

Fig. 6. Polarisation, $P$, measured for 25-μm cells under applied triangular-profile electric field of the amplitude 10V/μm and frequency 10Hz.

**SHG studies**

The polar nature of the mesophases was confirmed by second harmonic generation (SHG) experiments. SHG-microscopy measurements were performed on cooling from the isotropic phase for both materials in parallel rubbed 10 μm cells, with the fundamental wave (FW) polarization along the rubbing direction, i.e. the director, and with no analyser. The temperature dependence of the SHG signal (averaged over 80x80 pixels) is shown in Fig. 7(a). In both cases, the emergence of the SHG signal coincides with the transition between nematic phases. As an example, a comparison between polarising optical microscope and SHG-microscopy experiments is also shown for compound **NF8**. In polarising microscope, the transition between the non-polar and the polar nematic phases is marked by a texture destabilisation (Fig. 7(b)) and a progressive homogenization. Same sequence is observed in SHG-microscopy images. The initial sharp increase of SHG signal for both materials, is then followed by its stabilization. It is important to note here, that the signal depends on the refractive indices $(n_{2\omega} - n_{\omega})$, thickness of the cell (L) as well as on the effective nonlinear susceptibility $(d_{eff})$ being proportional to $d_{eff}^2 P_{\omega}^2 \sin^2\left(\frac{2\pi L}{\lambda}(n_{2\omega} - n_{\omega})\right)/(n_{2\omega} - n_{\omega})^2$, where $P_{\omega}$ is the power of the fundamental wave [50]. Thus, the evolution of the signal in Fig. 7(a) does not directly reflect the temperature dependence of nonlinear susceptibility coefficients.

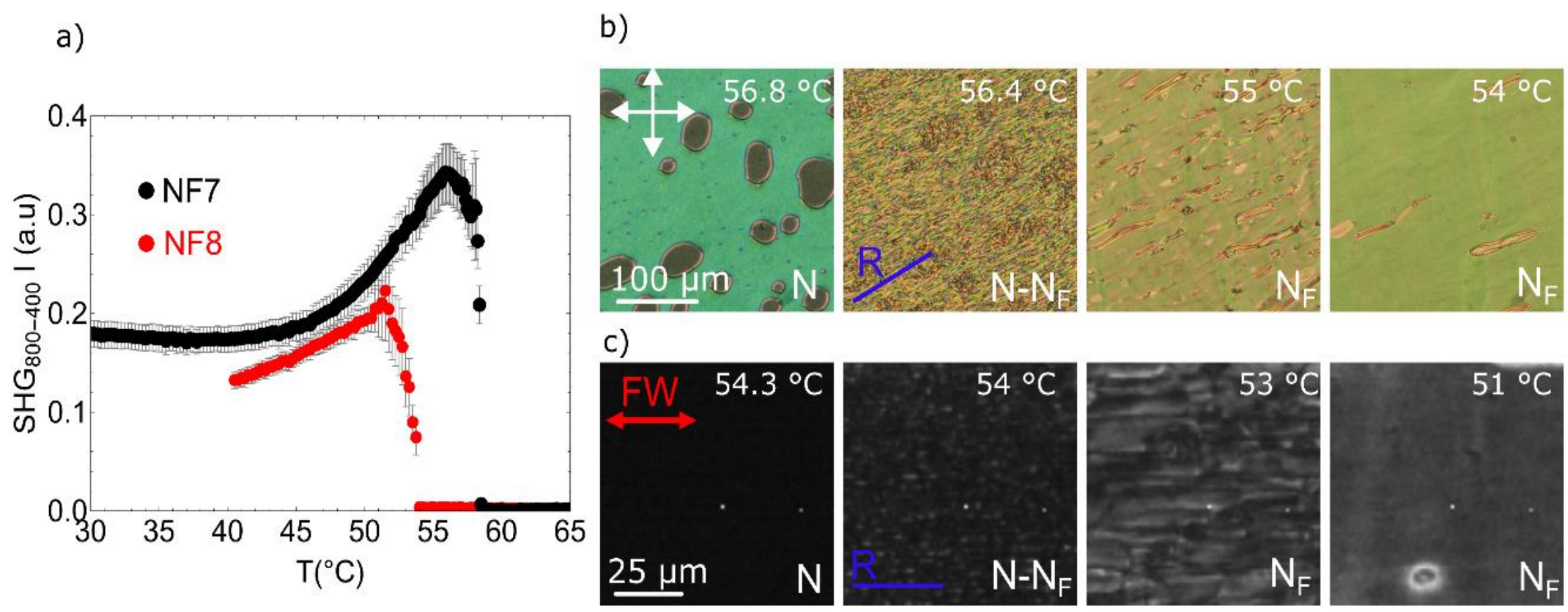

Fig. 7. (a) Temperature dependence of SHG signal recorded for **NF7** and **NF8** in 10- μm parallel rubbed cells on cooling. (b) Polarising optical microscopy images of the N-$N_F$ transition in **NF8**. Double-headed white arrows correspond to the crossed polarisers. Blue bar (R) corresponds to the rubbing direction. (c) SHG microscopy images for the same sample and the cell. Double headed red arrow corresponds to the polarisation direction of the incoming fundamental beam. Measurements are done without analyser.

To quantify the efficiency for SHG of the investigated compounds, the nonlinear optical coefficient $d_{33}$ was determined via Maker's Fringes method (see Experimental Section and SI). Incident light polarization and generated SHG light polarization were selected along the director and SHG intensity was recorded as a function of thickness. To ensure good director alignment, measurements were performed by applying a small DC voltage (5V for **NF7** and 10V for **NF8**). The $\sin^2\left(\frac{2\pi L}{\lambda}(n_e^{2\omega}-n_e^{\omega})\right)$ dependence allows for the determination of $\Delta n_d^{ee}=n_e^{2\omega}-n_e^{\omega}$, from the periodicity and $d_{33}$ from the maxima of the curve after setup calibration (See SI). Measurements were performed 2 degrees below the transition into the ferroelectric nematic phase. Unfortunately, sample crystallization during measurements prevented reliable determination of $d_{33}$ coefficients at lower temperatures. Fig. 8 shows SHG intensity as a function of thickness for both materials and the corresponding fits to $A\sin^2\left(\frac{2\pi L}{\lambda}(n_e^{2\omega}-n_e^{\omega})\right)$, which yield $\Delta n_d^{ee}=0.0309\pm0.0003$ and $d_{33}=6.8\pm0.7\ pmV^{-1}$ for **NF8** and $\Delta n_d^{ee}=0.0382\pm0.0002$ and $d_{33}=12.1\pm0.6\,pmV^{-1}$ for **NF7**. These values are among the largest reported for $N_F$ materials [24,51,52], especially considering that the measurements were performed at significantly longer wavelengths than in previous reports, where resonance enhancement typically leads to larger nonlinear coefficients.

We evaluated hyperpolarizability values of **NF7** through Gaussian calculations. Molecular geometry was optimized using DFT at the B3LYP/6-31G(d) level of theory and the obtained vibrational frequencies were checked in order to confirm that the final structure corresponds to an energy minimum. The hyperpolarizability tensor was then calculated within coupled perturbed Hartree-Fock (CPHF) theory. The main hyperpolarizability component is $\beta_{zzz}$ is along the dipole direction (which is almost completely parallel to the molecular long axis) and has a value of $41.96\times10^{-30}$ esu. This value allows us to estimate a theoretical upper bound for the main nonlinear susceptibility coefficient $d_{33}=Nf^3\beta_{zzz}$, where $N$ is the number density of molecules, $f = (n^2 + 2)/3$ is the Lorentz factor ($n$ being an average refractive index, taken here as 1.5), and perfect polar order has been assumed ($\langle \cos^3\theta \rangle = 1$) [51,24]. This calculation yields an estimated value of $d_{33} \approx 40$ pm V$^{-1}$. This value is approximately a factor of 4 larger than the one experimentally measured for **NF7**. However, one needs to consider that measurements were done just 2 degrees below the N-$N_F$ transition, so neither nematic nor polar order are saturated (see Fig. 7).

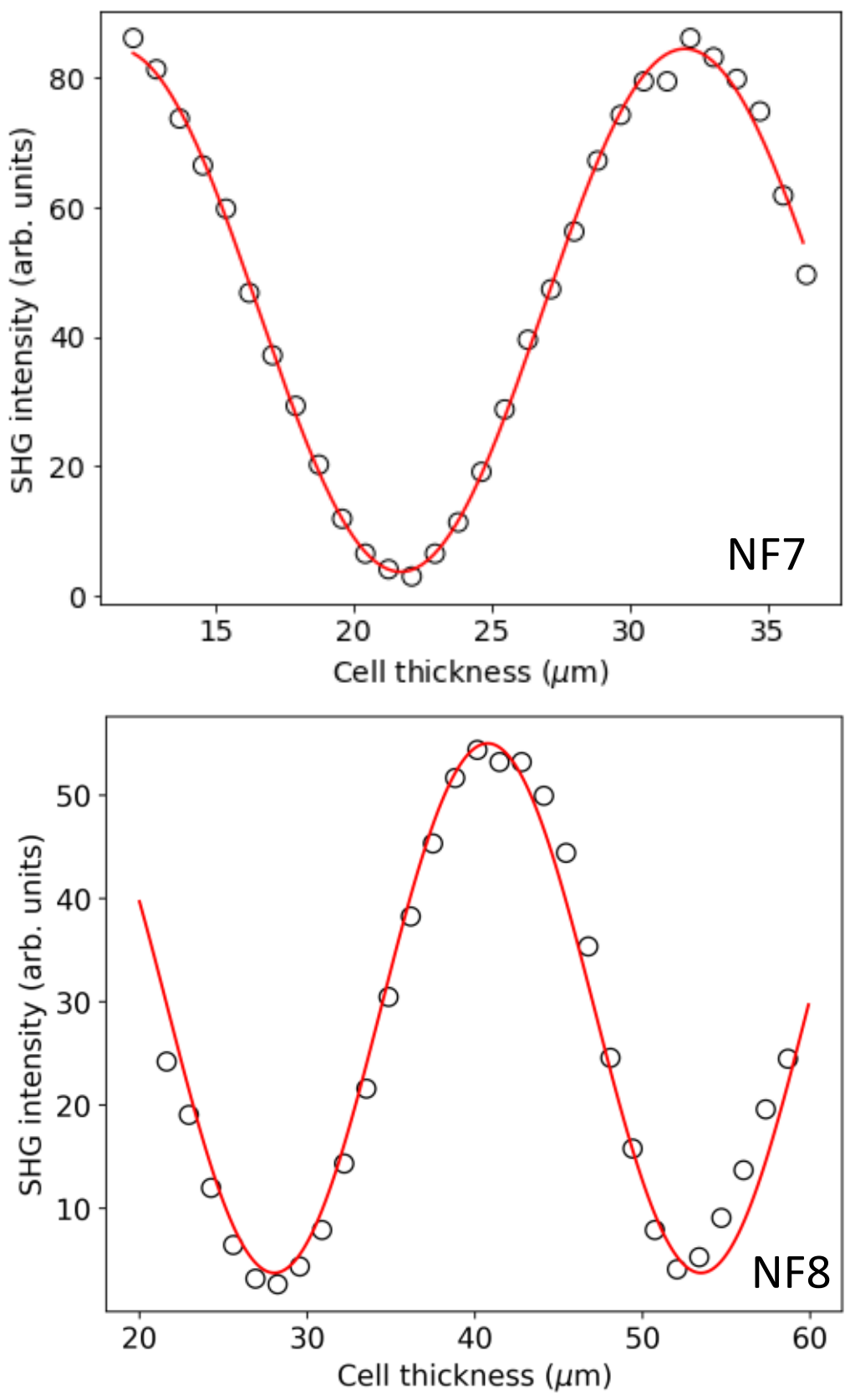

Fig. 8. SHG intensity vs. wedge cell thickness for (a) **NF7** and (b) **NF8**. Red lines correspond to fits to $A \sin^2\left(\frac{2\pi L}{\lambda}(n_e^{2\omega} - n_e^{\omega})\right)$.

**Birefringence measurements**

Optical birefringence was calculated using Eq. (1) from the retardation, which was measured starting from the isotropic phase on cooling through the temperature region of N and $N_F$ phases for **NF7** and **NF8** compounds. Contrary to the previously studied shorter homologues n=1-6, for **NF7** and **NF8** materials, due to the presence of the nematic phase, we managed to achieve perfect alignment in thin 1.6- μm parallelly rubbed cells (see Fig. 9(a) and (c) for **NF7** and **NF8**, respectively). For both compounds, nematic phase appearing below Iso was characterised with low-birefringent homogeneous textures, which was in consequence with the measured birefringence values (up to 0.015 for **NF7** and up to 0.01 for **NF8**), see Fig. 9(b). At the N- $N_F$ phase transition, a texture change front was observed (middle photos in Fig. 9(a,c)), which reflects molecular reorientation typical for this type of phase transition as polar intermolecular interactions and polar interaction with the surfaces start playing an important role. Simultaneously, an abrupt increase in optical birefringence was measured. For **NF8**, we observed a local minimum followed by further increase in birefringence values, which might be explained with strong fluctuations and molecular reorganisation (Fig. 9(b)). Subsequently, a gradual increase in birefringence was observed throughout the whole range of the $N_F$ phase, reaching the values 0.22 for **NF7** and 0.18 for **NF8** compounds. The difference in birefringence values obtained for these two compounds can be explained to be due to the next reasons. Birefringence of the material depends on the difference between the refractive indexes along the molecular axis and perpendicular to it, and hence on optical indicatrix. Due to the different length of the side chains, which influences the molecular packing and degree of nematic order, optical indicatrix of these two materials could differ as a result of different nematic order parameters. We suggest that due to the disturbing influence of the side chain on the molecular ordering, longer NF8 material possesses smaller nematic order in comparison with shorter homologue NF7, thus lower birefringence can be expected.

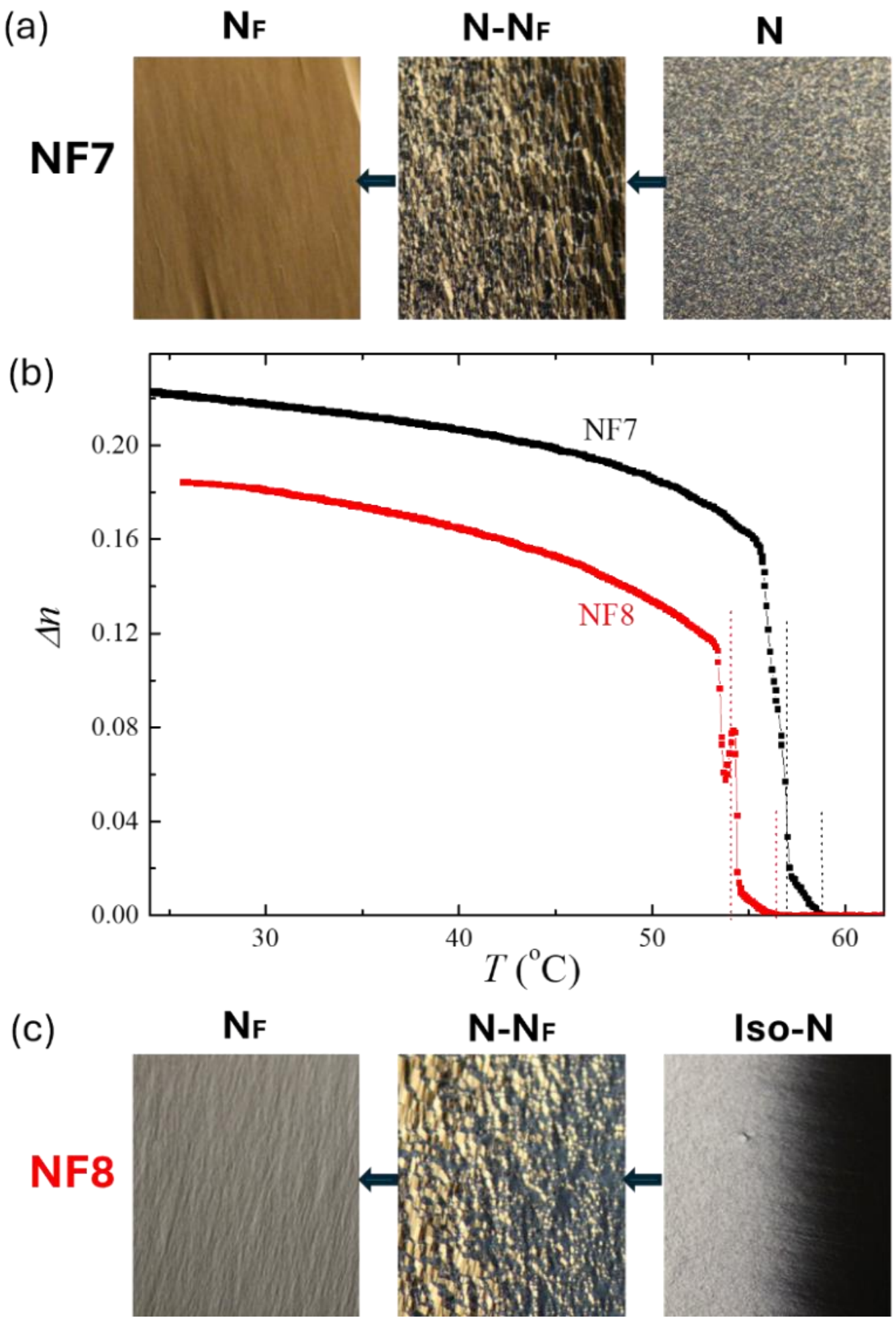

Fig. 9. (a) Textures in the nematic, nematic to ferroelectric nematic phase transition, and ferroelectric nematic phases for **NF7** material; (b) optical birefringence in dependence on temperature for **NF7** (black symbols) and **NF8** (red symbols); (c) textures at the isotropic to nematic, nematic to ferroelectric nematic and in the ferroelectric nematic phase for **NF8** compound. The width of the photos is 400 μm. Crossed polariser and analyser are along the edges of the photos.

## Discussion and conclusions

In our previous research, we introduced a molecular structure with dimethylamino-terminated terminal part as electron-donating group [37]. Contrary to the previously studied compounds **NFn** with a shorter side-chain length up to n=6, new compounds **NF7** and **NF8**, presented in this work, exhibit a narrow nematic phase above the ferroelectric nematic one. Detailed calorimetric analysis confirmed the presence of two mesophases, N and $N_F$, by the appearance of two distinct peaks in the temperature dependence of the heat capacity. Additionally, specific behaviour of the N-$N_F$ phase sequence is demonstrated in textures observed under a polarized light microscope.

Observation of the ordinary nematic phase with prolongation of the side alkoxy-chain seems contradictory to the trend described by Cruickshank [53]. For methoxy-terminated RM734/benzoate type of materials, the prolongation of the chain changes the aspect ratio of the molecule and destabilizes ordinary nematic phase more significantly than the polar nematic phase. Suggested origin is that the *anti*-parallel (head-to-tail) arrangement of the molecules in the ordinary N phase is disfavoured by the presence of the bulky lateral chains. On the other hand, when the molecules are oriented *syn*-parallel but slanted as typical for polar nematic phase, it causes less hindrance. The effect is even more pronounced when the alkoxy chain is located at the central phenyl ring [54], which seems natural as it facilitates the favourable intermolecular overlap of the first and the third aromatic ring with complementary electron density. Authors [53,54] propose that longer alkyl chains tend to adapt more parallel arrangement, which results in preference of ferroelectric nematic phase.

In a similar way, Huang et al [55] suggest comprehensive approach to explain the tendencies for polar or apolar N phase, using a set of molecular parameters describing shape anisotropy and polarization. The main advantage of such approach is quantification of the multiple molecular parameters, which supports the above-mentioned destabilisation effect on apolar N phase. Authors of [55] also concluded that the incidence of apolar N phase between polar N and isotropic liquid is given mostly by the shape anisotropy factor and to less extend by the magnitude of dipole or dipolar density, which agreed with vast majority of analysed polar nematic materials.

Nevertheless, we calculated the parameters of our materials (see SI Fig S12 for comparison of transition temperature trends and Table S1) and found that our compounds do not correlate with above mentioned conclusions. There are several key properties of our materials: very high dipole moment (ca. 14 D) almost parallel to the long molecular axis (deviation 9°); high dipolar density parameters; high anisotropy factors. Combination of all those parameters suggests strong intermolecular interactions in such molecular system. Generally, dimethylamino is a better electron-donating group than methoxy and provides higher electron density, polarizability and resulting higher dipole moment. We can speculate, that if the attractive dipolar forces in NFn materials are stronger than in methoxy analogues, it might be possible for molecules to adapt also antiparallel organisation of the apolar N phase despite the steric hindrance caused by lateral alkyl chains. Additionally, the dimethylamino group is bulkier than methoxy so elongation of the alkyls for NFn materials can play different role. Change of methoxy terminal group to ethoxy led to significant drop of $T_{iso}$ [10], but a change from dimethylamino to diethylamino led to total suppression of LC behaviour [56].

In the $N_F$ phase in homogeneous cell with antiparallelly rubbed surface, we observed usual tendency to create twisted domains. However, besides twisted domains, another type of domains of elongated shape and with borderlines oriented along the rubbing direction were observed. We can speculate about different molecular orientation within these domains throughout the cell bulk. The origin and the structure of such domains will be presented in a separate study.

We concentrated on the physical properties and parameters of the ferroelectric $N_F$ phase with respect to the molecular structure. We studied the polar properties and confirmed the ferroelectric character of the $N_F$ phase and the weakly first-order character of the N-$N_F$ phase transition. We have established dielectric properties and for compound **NF7**, we compared the results of dielectric measurements in similar cells, differing only in the presence of surfactant layers and analysed the sample in a commercial 7-μm HG-A cell and a home-made cell, prepared without surfactant layers. Interpretation of the dielectric spectroscopy data is a rather controversial subject in the $N_F$ phase [45-49]. The results of dielectric spectroscopy are to a great extent influenced by the surface effects at the cells boundaries and thus it is not straightforward to evaluate the material properties solely from them. The use of aligning layers may cause undesired charge accumulation [47]. Surface layers serve as an additional large capacitance in series with bulk LC and thus the dielectric spectroscopy data interpretation must account for these complications. Even in the absence of surfactant, one must be careful as thin LC layers close to the surface, in which the polarisation direction changes from the one preferred by the surface to the one preferred by the bulk, serve similarly to the alignment layers [47]. PCG model proposed by Clark et al [45] is an effective model for the evaluation of the dielectric properties of $N_F$ materials. In this model, due to high polarisation values in the $N_F$ phase, the resistivity of these materials was considered to be low, as it is inversely proportional to the square of polarisation magnitude. Nevertheless, as the resistivity is proportional to the rotational viscosity, its influence should not be neglected [48]. The viscosity of $N_F$ materials has been predicted to be high [48,57], and it is certainly confirmed to be largely exceeding the viscosity of classical nematics, especially at ambient temperatures [58].

To summarize, we have demonstrated that the combination of dimethylamino moiety (the strongest neutral electron donating group) and the nitro moiety (an effective electron withdrawing group) at the opposite end of the molecule, connected by a $\pi$-conjugated bridge, makes **NF7** and **NF8** materials an efficient push-pull system with extensive electron delocalisation. Such delocalization of electrons along the long molecular axis is reflected in the molecular hyperpolarizability values and experimentally shown by the large nonlinear optical $d_{33}$ coefficients measured for fundamental wave in the telecommunication band at 1.574 μm.

We proved that contrary to previously reported data for other materials, for studied series NFn the prolongation of the side chain leads to an appearance of the nematic phase above the $N_F$ mesophase. These results highlight the importance of designing molecular architectures that balance the intermolecular interactions required for the stability of the polar $N_F$ phase and the addition of effective push-pull systems, attractive for nonlinear photonic applications operating in the silicon photonics wavelength range.

**Author contributions**

N.P.: materials characterisation (sample preparation, microscopic observations, dielectric spectroscopy fitting and evaluation of the results, birefringence measurements), manuscript writing; M.C.: synthesis, chemical characterisation and writing of the corresponding parts of the manuscript; P.G.: cells preparation, DSC, dielectric spectroscopy and polarisation measurements; M.L. and G.C.: High resolution AC calorimetry measurements and writing of the corresponding parts of the manuscript; A.S.: custom wedge cell preparation; A.S., N.S & A.M.: SHG microscopy; A.E., J.M.-P and N.S: NLO coefficients measurements and analysis; N.S and A.E writing of the corresponding parts of the manuscript; V.N.: microscopic observations, dielectric and polarisation measurements evaluation, curation of the research, manuscript writing.

**Conflict of interest**

The authors declare no conflict of interest.

**Funding**

Authors acknowledge the financial support of the Czech Science Foundation, the project 24-10247K, and the Slovenian Research and Innovation Agency (grant numbers P1-0192, P1-0125, L1-70060, and N2-0367). A.E. and J. M.-P. acknowledge funding from the Basque Government Project IT1979-26 and from project PID2023-150255NB-I00 from MCIU/AEI/10.13039/ 5011000-11033/FEDER, UE.